\documentstyle[12pt]{article}

\newskip\humongous \humongous=0pt plus 1000pt minus 1000pt

\newif\ifdtup

\catcode`\@=11

\@addtoreset{equation}{section}
\def\theequation{\thesection\arabic{equation}}

\def\@normalsize{\@setsize\normalsize{15pt}\xiipt\@xiipt
\abovedisplayskip 14pt plus3pt minus3pt%
\belowdisplayskip \abovedisplayskip
\abovedisplayshortskip \z@ plus3pt%
\belowdisplayshortskip 7pt plus3.5pt minus0pt}

\def\small{\@setsize\small{13.6pt}\xipt\@xipt
\abovedisplayskip 13pt plus3pt minus3pt%
\belowdisplayskip \abovedisplayskip
\abovedisplayshortskip \z@ plus3pt%
\belowdisplayshortskip 7pt plus3.5pt minus0pt
\def\@listi{\parsep 4.5pt plus 2pt minus 1pt
     \itemsep \parsep
     \topsep 9pt plus 3pt minus 3pt}}

\relax

\catcode`@=12

\catcode`\@=11

\def\section{\@startsection{section}{1}{\z@}{3.5ex plus 1ex minus
   .2ex}{2.3ex plus .2ex}{\large\bf}}

\def\thesection{\arabic{section}.}

\def\appendix{\setcounter{section}{0}
 \def\thesection{Appendix \Alph{section}:}
 \def\theequation{\Alph{section}.\arabic{equation}}}

\begin{document}

\begin{titlepage}
\begin{center}
{\Large
 Fermions in
Instanton-Anti-Instanton Background }
\end{center}

\vspace{1em}
\begin{center}
{\large
Riccardo Guida and Kenichi Konishi}
\end{center}

\vspace{1em}
\begin{center}
{\it Dipartimento di Fisica -- Universit\`a di Genova\\
     Istituto Nazionale di Fisica Nucleare -- sez. di Genova\\
     Via Dodecaneso, 33 -- 16146 Genova (Italy)\\
     E-mail: Decnet 32655; Bitnet @GENOVA.INFN.IT\\}
\end{center}

\vspace{7em}
{\bf ABSTRACT:}
We consider the behaviour of fermions
in the background
of instanton-anti\-instanton type configurations.
Several different physics problems,
from the high energy electroweak interactions to
the study of vacuum structure of QCD and of large orders
of perturbation theory are related to this
problem.
The spectrum of the Dirac operator in such a background is studied in detail.
We present an approximation for the fermion correlation function when
the instanton-anti\-instanton separation ($R$) is large compared to their
sizes ($\rho $).
The situation  when
 the  instanton-anti\-instanton overlap and melt, is studied
through the behaviour of the Chern Simons number as a function of $ R/\rho$
and $x_4$.
 Applying our results to widely
discussed cases of fermion-number violation in the electroweak theory,
we  conclude that there are  no theoretical basis
for expecting  anomalous cross sections to become observable at
energies in  $10$ TeV region.
\vspace{2em}
\begin{flushleft}
GEF-Th-8/1993~~~~~~~~~~~~~~~~~~~~~~~~~~~~~~~~~~~~~~~~~~~~~~~~~~~~~~~~
{}~~~~~~~~~ March 1993 \end{flushleft}
\end{titlepage}

\newcommand{\beq}{\begin{equation}}
\newcommand{\eeq}{\end{equation}}
\newcommand{\bea}{\begin{eqnarray}}
\newcommand{\eea}{\end{eqnarray}}
\newcommand{\beas}{\begin{eqnarray*}}
\newcommand{\eeas}{\end{eqnarray*}}
\newcommand{\defi}{\stackrel{\rm def}{=}}
\newcommand{\non}{\nonumber}

\def\dinv{{\bar D}^{-1}}
\def\et0{\eta^{(a)}_0}
\def\emi{\eta^{(i)}_m}
\def\zema{{\bar \zeta}^{(a)}_m}
\def\etm{\eta^{(a)}_m}
\def\etn{\eta^{(a)}_n}
\def\zet0{{\bar \zeta}^{(i)}_0}
\def\zetn{{\bar \zeta}^{(i)}_n}
\def\zetm{{\bar \zeta}^{(i)}_m}
\def\dainv{({\bar D}^{(a)})^{-1}}
\def\cbar{{\bar C}}
\def\bbar{{\bar B}}
\def\d00{{\bar D}_{00}}
\def\dbar{{\bar D}}
\def\dabar{{\bar D}^{(a)}}
\def\dibar{{\bar D}^{(i)}}
\def\proja{{\bf 1}-|a,0 \rangle\langle a,0|}
\def\proji{{\bf 1}-|i,0\rangle \langlei,0|}
\def\bra{\langle}
\def\ket{\rangle}
\def\sbar{\bar S}

\def\dirac{{\cal D}}
\def\dplus{{\cal D_{+}}}
\def\dminus{{\cal D_{-}}}
\def\de{\partial}
\def\si{\sigma}
\def\sb{{\bar \sigma}}
\def\rn{{\bf R}^n}
\def\r4{{\bf R}^4}
\def\s4{{\bf S}^4}
\def\ker{\hbox{\rm ker}}
\def\dim{\hbox{\rm dim}}
\def\sup{\hbox{\rm sup}}
\def\inf{\hbox{\rm inf}}
\def\infi{\infty}
\def\nrm{\parallel}
\def\nrmi{\parallel_\infty}
\def\teo{\noindent{\bf Theorem}\ }
\def\tt{\tilde T}
\def\st{\tilde S}
\def\om{\Omega}
\def\sprime{S^{\prime}_{x,y} }
\def\i-a{instanton-anti-instanton}

\def\qi{{\cal Q}_i}
\def\calpi{{\cal P}_i}
\def\qa{{\cal Q}_a}
\def\pa{{\cal P}_a}
\def\calf{{\cal F}}
\def\calg{{\cal G}}

\section{Introduction.}
The purpose of the present paper is to study the behaviour of chiral
fermions in the background of instanton-anti-instanton type. In particular
we wish to understand how unitarity works out in the presence
 of topologically nontrivial
effects such as instantons, treated within the
semiclassical approximation.

 The interest
in this problem   arises at least from three different
sources.

 First,
there does not seem to be a universal consensus as yet among the
physicists  whether
 the fermion-number violation becomes strong in the TeV region
scatterings, induced by the instanton or  sphaleron in the SU(2)
electroweak interactions [1-30].
 Both direct calculations of the cross section by
the instanton method as in the works of \cite{3,8}, and another approach
\cite{6,15,22}
which makes use of the optical theorem and the so-called valley method,
encounter various technical difficulties. The opinion in favour of the
anomalous processes becoming observable, has been expressed based on some
toy-model calculations \cite{15}, or on some general considerations \cite{26}.
In the valley approach the question of unitarity and the  behaviour of the
fermion propagators in the background of instanton-anti-instanton background
are quite central.
The results of our investigation show that there are no theoretical basis
in the claims made in the literature that the anomalous cross sections
become observable  in the TeV region scattering in the standard electroweak
model.\footnote{There are also  arguments
\cite{9,10,14,17,18,bachas}, which rely on the
 unitarity constraints in  multiple gauge-boson productions, that such
processes remain necessarily suppressed by a finite fraction of the 't Hooft
factor,
$ \sim \exp (-4 \pi / \alpha).   $
 As far as the authors know these arguments have never been rebuked,
but at the moment it seems to be difficult to make them more quantitative
and   rigorous.}

Secondly our  study is closely related to that
of the vacuum structure in QCD, especially in connection with the so-called
instanton liquid model \cite{37}, although in the latter
one needs the knowledge of
fermion propagators in the multi-instanton background. A detailed study of
the fermion propagation in the \i-a background  should help understanding
of a  more complex situation in the multi-instanton media.

Finally the contribution of \i-a to the correlation functions has a deep
connection with the problem of large order perturbations both in QCD and in
the electroweak theory. Many related issues (Borel summability,
determination of the large order coefficients for R, the role of renormalons
, etc.)  are receiving a renewed interest in the literature \cite{lar,bachas}.
\smallskip

To be definite we consider an $SU(2)$ gauge theory with $ N_F $  left\--handed
fermi\-on doublets.

The main problem regarding unitarity and chiral anomaly
can be formulated as follows.
\footnote{The work of Ref.\cite{11} (see also
Ref.\cite{24}) goes quite
some way
in proving unitarity in the presence of instanton effects,
 especially as regards
 the final states (i.e., states summed over in the
unitarity relations).  However,  fermions are not
considered there: as a result no discussion on subtleties related to
chiral anomaly is found in \cite{11}.  In any case, the  central issue related
with the initial (or external) states
has not been addressed before. }

 The optical theorem states that the cross section,
\beq 1 + 2 \longrightarrow X, \label{1_1} \eeq
summed over $X$, is equal, apart from a kinematical factor, to the imaginary
part of the forward elastic amplitude,
\beq 1 + 2 \longrightarrow 1 + 2. \label{1_2} \eeq
Now consider a particular class of processes (\ref{1_1})
induced by an instanton ("anomalous processes"),
with the change of the fermion number,
\beq\Delta f = f_1+f_2 -f_X = N_F. \label{1_3}\eeq
Sum over the final states satisfying (\ref{1_3}) should
 give a contribution to  the imaginary
part of the elastic amplitude.

For an (anti-) instanton background, which is relevant for the calculation
of the production process (\ref{1_1}), each right (left) handed fermion field
has a zero mode. The standard functional integration over fermions yields  a
product of these zero modes; by going to momentum space and by applying
the LSZ amputation one finds the S-matrix elements consistent with (\ref{1_3}).

The corresponding contribution in the elastic amplitude (\ref{1_2}) must
arise from
a sort of instanton-antiinstanton ($i-a$)
 background, topologically (globally)
equivalent to the trivial, perturbative vacuum.  One  expects however
no fermion zero modes to exist in such a background (see Section 2 for more
details). How  can one compute
the "anomalous" part of the elastic amplitude then?

On a general physical ground one expects that at large $i-a$ separation
($R$) the standard fermion zero modes in the anti\-instanton or the instanton
background should play an important role.  For instance the two point
function,
\beq I(x,y) = \int {\cal D}\psi{\cal D}{\bar \psi}\, \psi(x) {\bar \psi}(y)
\exp -\!\int \! d^4x\, i\,{\bar \psi}{\bar D}\psi,\label{1_4} \eeq
is expected to behave approximately as:
\beq I(x,y) \simeq \et0\!(x)\,\zet0\!(y)^*  \label{1_5}\eeq
where $ \et0\!(x) $  ($ \zet0\!(y) $) is the left-handed (right-handed)
zero mode in the anti\-instanton (instanton) background. Such a
behaviour was simply assumed in the existing literature so far.
    We wish  however
 to {\it compute} $I(x,y)$, prove that Eq.(\ref{1_5}) indeed holds
approximately at
large $R$, and to calculate
the corrections.
  To do this requires a systematic way of calculation.

To make the problem well-defined,
we consider
a particular
class of
$i-a$ type configurations - the so called valley, or streamline, trajectory
\cite{30,31,15}
\beq  A_{\mu}^{(valley)} =  A_{\mu}^{(a)} +  A_{\mu}^{(i)} +   A_{\mu}^{(int.
)}, \label{1_6}  \eeq
\beas
A_{\mu}^{(a)}   & =& -{i\over g}(\sigma_{\mu} {\bar \sigma_{\nu}}
-\delta_{\mu \nu} ) {(x-x_a)_{\nu} \over (x-x_a)^2 +\rho^2 }, \non\\
A_{\mu}^{(i)}   & =& -{i\over g}(\sigma_{\mu} {\bar \sigma_{\nu}}
-\delta_{\mu \nu} )
{(x-x_i)_{\nu} \rho^2 \over (x-x_i)^2 ((x-x_i)^2 +\rho^2) }, \\
A_{\mu}^{(int.)}   & =& -{i\over g}(\sigma_{\mu} {\bar \sigma_{\nu}}
-\delta_{\mu \nu} )
 [ {(x-x_i +y)_{\nu} \over
(x-x_i + y )^2  } -  {(x-x_i)_{\nu} \over
(x-x_i)^2  }],
\eeas
where
\beq
 y = -R/(z-1);\,\,
 z = (R^2 + 2 \rho^2 + \sqrt{R^4 + 4\rho^2 R^2}) /2\rho^2;\,\,
R^{\mu} = (x_i - x_a)^{\mu}. \label{1_7} \eeq

There are several reasons for such a choice. First, the classical field
Eq.(\ref{1_6}) is known, at least at large $R$, to have the correct form of
 interaction between the instanton and anti\-instanton (dipole-dipole).
Secondly, Eq.(\ref{1_6})
 interpolates two solutions of the Euclidean field equations,
$A_{\mu} = A^{(i)}_{\mu} +A^{(a)}_{\mu} $ (at $R=|x_a-x_i| = \infty$)
  and   (gauge-equivalent of)
$ A_{\mu}=0 $  (at $R =0$ ), so that the phenomenon of $i-a$ "melting"
at $ R \rightarrow 0$  can be
studied quantitatively.
Thirdly, they are  solutions  of the valley equation \cite{15}.
\footnote{This is true only in  unbroken gauge theories
such as QCD. In the presence of Higgs fields phenomenological relevance of
the valley configuration Eq.(\ref{1_6}) is not  obvious.  We
concentrate here on the problem of unitarity versus anomaly as formulated
above, in the (relatively) simple setting of Eq.(\ref{1_6}), which is the
starting
point also for similar problems in the standard electroweak theory. }
Importance of the  non-Gaussian integrations
along such an almost flat valley in the field configuration space, was
emphasised first in Ref.\cite{30,31} in a general context of
quantum mechanics and  QCD.
\footnote{We believe that,
for a well-separated
instanton-anti\-in\-stan\-ton pair,
 the val\-ley method   does give the
dominant contribution to the anomalous imaginary
part of amplitudes. For ${R\over \rho}\le 1$, instead,
we see no reasons to expect such a valley field to be physically
distinguishable from generic perturbations around $A_\mu =0$ (see Section
[4.1]), hence to be of particular importance in the functional integration.
Nonetheless the use of a concrete and explicit valley field such as
Eq.(\protect\ref{1_6}) is quite adequate for our purposes. }
Furthermore, the valley field of Eq.(\ref{1_6}) satisfies the simple, covariant
gauge condition,
\beq \partial_{\mu}  A_{\mu}^{(valley)} = 0, \label{1_8} \eeq
so that all calculations can be done in a manifestly covariant fashion.

\smallskip
We shall study  the four point function,
$$<T\psi_1(x) \psi_2(u){\bar \psi}_1(y) {\bar \psi}_2(v)>^{(A_{valley})}
$$
\beq =\int{\cal D}\psi{\cal D}{\bar \psi} \, \psi_1(x) \psi_2(u)
 {\bar \psi}_1(y) {\bar \psi}_2(v)
\,{\rm e}^{-S} /{\cal Z}^{(A=0)}; \label{1_9}\eeq
$$S=\sum_{j=1}^{N_F} \int d^4x\, i\,{\bar \psi}_j {\bar D}\psi_j      $$
in the fixed background of Eq.(\ref{1_6}). Integrations over the collective
coordinates such as $R$ and $\rho$ are to be performed afterwards.

As the functional integral factorises in flavour we must study (suppressing
the flavour index),
\beq I(x,y) = \int {\cal D}\psi{\cal D}{\bar \psi}\, \psi(x) {\bar \psi}(y)
\exp -\!\int \! d^4x\, i\,{\bar \psi}{\bar D}\psi, \label{1_10} \eeq
and
\beq
{\cal Z}=  \int {\cal D}\psi{\cal D}{\bar \psi}
\exp -\!\int\! d^4x \,i\,{\bar \psi}{\bar D}\psi = \det \dbar, \label{1_11}
\eeq
where it is assumed that $\det \dbar$ is suitably regularised.

\smallskip
The paper will be  organised as follows. First we study in
 Section 2 (also in Appendix A and
Appendix C) the spectrum of the Dirac operator in the valley background
 Eq.(\ref{1_6})
in detail.  Among others, the absence of the fermion zero modes
(and actually  of any normalisable modes) is proved.
 In Section 3 the behaviour of the
fermion Green functions is studied at large $R/\rho$, leading
essentially to the
behaviour Eq.(\ref{1_5}).   In Section 4 we study
the situation at small $R/\rho$ where the instanton and anti-instanton
overlap on each other and melt. Although this analysis is somewhat
indirect, being
based on the behaviour of the Chern Simons number for the background
Eq.(\ref{1_6})  as a function of $R/\rho$
and of $x_4$, it leads to a semi-quantitative idea of when the \i-a pair
actually  melt.

\section{Spectrum of the Dirac Operator in the Valley
Background }
We first want to learn all we can about the   Dirac operator $\dirac$ in the
valley background and in the full euclidean space $\r4$, where:
$$
\dirac = \pmatrix {0&\dminus\cr
                   \dplus&0},$$
$$\dplus = i \dbar = i\sb _\mu D_\mu , \;\;\;\dminus= i D =i\si _\mu D_\mu$$
and $D_\mu=(\de-igA)_\mu$ is the usual covariant derivative. The discussion
will be divided in three parts, concerning the essential spectrum
\footnote{\rm The usual definition of {\bf essential spectrum } of an
operator (see \cite{RSA}) is
$\si_{ess}\equiv\si / \si _{disc}$, where the {\bf discrete spectrum }
 $\si _{disc}$
is the set of isolated point of the spectrum that are eigenvalues with
finite multiplicity.},  the zero modes, and finally the positive
(normalisable) modes.

Let us  consider the valley field
$$A_\mu =-{i\over g}\sb_\rho (\si _\mu \sb _\nu -\delta_{\mu \nu})\si _\tau
{ v_\rho v_\tau \over v^2}\,
 H_\nu $$
$$ H_\nu =
 {x-x_a\over (x-x_a)^2+\rho ^2}-{x-x_i\over (x-x_i)^2+\rho ^2}
\;\;\;\;\;\;( v=x-x_i+y )$$
which is obtained  from Eq.(\ref{1_6}) by the gauge transformation:
$\, U={\sb _\mu v_\mu \over \sqrt{v^2}}\, $   (which obviously does not change
 the spectrum of our operator).
The advantage of this gauge choice ("clever gauge")
 is  that $A_\mu(x)$  is now a bounded
function
\footnote{If we define  $\nrm f\nrmi\equiv
 \sup_x |f(x)|$ (i.e. the minimum upper bound of $|f(x)|$),
then $\nrm A_\mu \nrmi < \infi$. }
and that $A_\mu (x)\sim {R_\mu \over |x|^2}$ as $x \to \infi$, (which implies
the uniformity  of convergency in the direction of $x$.)
We show in Appendix A that these properties allow us to prove the theorem
$$\si (\dirac)\equiv \si _{ess} (\dirac)=(-\infi ,+ \infi ):$$
the essential spectrum is the same as that of  the free Dirac operator.

We now turn our attention to the possible zero modes of the Dirac
operator in the valley background.
Thanks to the property $\int _{\r4} F^2<\infi $ of our vector potential,
we can consider (see \cite{NIELS}) the Dirac operator trasported
by stereographic mapping
 in the compactified space
$\s4 $ (which we call $\hat{\dirac}$),
and apply  the  Atiyah-Singer theorem \cite{as} to evaluate the index of
differential operators in compact manifolds to $\hat{\dirac}$.
This yields  the well known result :
$$ \hat{N}_+-\hat{N}_- =-{g^2\over 16\pi}
\int_{\r4 } Tr(F\tilde{F})$$
where $ \hat{N}_{\pm}=\dim( \ker \hat{\dirac} _{\pm})$, i.e the
number of normalisable (in the sense of $\s4$ measure) zero modes of the
operator with left (resp. right)
 helicity. It can be proved that zero modes $\hat{\psi}$ of $\hat{\dirac}$
can be mapped into zero modes $\psi$ of $\dirac$ (and viceversa)
and that the normalisation condition for $\hat{\psi}$ in $\s4$
 gives the condition
$$\int d^4\! x (1+x^2)^{-1} |\psi(x)|^2<\infi , $$
 (i.e. they are in $L^2(\r4, {d^4\!x \over (1+x^2)})$), which does not
guarantees the usual normalisation condition. Clearly a zero mode of $\dirac$
normalisable in the usual sense can be mapped in a normalisable zero mode
of $\hat{\dirac}$, so that we have $N_{\pm}\leq  \hat{N}_{\pm}$,
where $ N_{\pm}=\dim( \ker \dirac _{\pm})$, i.e the
number of normalisable zero modes of euclidian
 Dirac operator with left (resp. right)
 helicity.
In the valley background
$\int_{\r4  }Tr(F\tilde{F})=0$.  The question is:  {\it  is "0=0-0" or
  "0=1-1" or worse ?,
i.e,  are there  zero modes in the valley background ?}

Because the valley is essentially the sum of istanton and
 anti-istanton background
(each with one zero mode, but with opposite helicity)
 one  expects that $\dirac$  can have at most
two zero modes with different helicity.
    The following assertion will now be proved:

\noindent{\bf  "$0\neq 1-1$. "}
\footnote{The claim is valid also for zero modes not normalisable
in the usual sense, but that are in $L^2(\r4, {d^4\!x \over (1+x^2)})$, it
suffices to modify the normalisation condition, without other changes.}

\noindent{\bf Proof:}
Let us go back to Lorentz gauge of Eq.(\ref{1_6}):
\beq A_\mu =-{i\over g} (\si _\mu \sb _\nu -\delta_{\mu \nu})
\, F_\nu = \bar{\eta}_{\mu\nu}^a F_\nu \si ^a\eeq
\beq
F_\mu(x)\equiv {1\over 2}\,\de_\mu\log L(x)\;\;\;\;\;\;
L(x) \equiv {(x-x_a)^2+\rho^2\over (x-x_i)^2+\rho^2}\,(x-x_i+y)^2.
\label{elle}\eeq
We assume (without less of generality) that
$x_i=(-{R\over 2},0,0,0)$ and $x_a=({R\over 2},0,0,0)$, so that also
the parameter $y$ above has the form
$y=(y,0,0,0)$ (see appendix D for definition of $y$).

It turns out to be simpler for our purposes  to work in
 a noncovariant formalism:
$$
\dminus=-\de_t +i\vec{\si}\cdot\vec{\nabla} +i\vec{\tau}\cdot\vec{F}
-\vec{\si}\wedge \vec{\tau}\cdot\vec{F}-F_0 \vec{\si}\cdot\vec{\tau}$$
$$
\dplus=+\de_t +i\vec{\si}\cdot\vec{\nabla} -i\vec{\tau}\cdot\vec{F}
-\vec{\si}\wedge \vec{\tau}\cdot\vec{F}-F_0 \vec{\si}\cdot\vec{\tau}$$
 where $x_\mu\equiv(t,\vec{r})$,
$$\vec{F}=\vec{r}f(r,t);\;\;\; F_0=F_0(r,t)$$
$r=\sqrt{{\vec{r}}^2}$,  $\vec{\si} $ are spin operators, and
$\vec{\tau } $ are isospin one.
First let us  note that $\dirac$ commutes with the threedimensional
total angular momentum operator
$\vec{J}=-i\vec{r}\wedge\vec{\nabla} +\vec{\si}+\vec{\tau},$
so that any eventual normalisable eigenmode of $\dirac$
 can be chosen to
be also eigenmode of $J^2$ and $J_3$, with quantum number $j,m$, as usual.
Note that the symmetry forces eigemodes to be in $2j+1$-plets,
so if we find a zero mode with total spin $j$ then there must be
 $2j$ more zero eigenmodes.  Since
we are interested in  a single zero mode each of $\dplus$ and
 $\dminus$ for the reasons mentioned above,
we look for a solution with  $j=0$.
Let us concentrate on the lefthanded mode, associated with $\dplus$
below.

There are  two ways to form
a $j=0$ state, because the fermion wave function is in a
doublet representation both of isospin and of spin.
Thus,  $\si$ and $\tau$  combine
into  a singlet or triplet representation,  coupled
 with a singlet
or  triplet  representation  respectively
of spatial angular momentum,  to form
a singlet of the total spin.
The most general function with $j=0$ can  be written accordingly  as:
$$\eta_{j\alpha}=
(\si_2)_{j\alpha}S(r,t)-i(\vec{\si}\si_2)_{j\alpha}\cdot\vec{x}\,T(r,t).$$
The action of $\dplus$ on $\eta$ is:
$$\dplus\eta=
\si_2( \de_tS+3F_0S- \vec{x}\cdot\vec{\nabla} T-3T+\vec{F}\cdot\vec{x}\, T)
+i\vec{\si}\si_2\cdot
(-\vec{\nabla} S-3 \vec F S- \vec x \de_tT+\vec  x F_0T).$$
Imposing $\dplus\eta=0$, we get the system of equations:
\bea
\de_tS+3F_0S- \vec{x}\cdot\vec{\nabla} T-3T+\vec{F}\cdot\vec{x} T
&=&0,\non\\
-\vec{\nabla} S-3 \vec F S- \vec x \de_tT+\vec  x F_0T&=&0\label{2_1}.\eea
Eq.(\ref{2_1}) can be simplified by defining
 two new functions $\tt (r,t)$ and $\st (r,t)$:
 $$T={1\over r^3}{L}^{1\over 2} \tilde{T};
\;\;\;S=L^{-{3\over 2}} \tilde{S},$$
($L(x)$ is defined in (\ref{elle})), so that Eq.(\ref{2_1}) reads:
\bea \de_t \st - h\de_r \tt&=0;\non\\
-\de_r \st - h\de_t \tt&=0;\label{2_2}\eea
where
 $$h=h(r,t)={L(r,t)^2\over r^2}.$$
The consistency condition $\,\de_r\de_t\st=\de_t\de_r\st\,$
following from  Eq.(\ref{2_2}) is equivalent
to $\nabla (h \nabla \tt)=0$ (here $\nabla\equiv (\de_r,\de_t)$).
Furthermore
the normalisation condition $\int |\eta|^2<\infi $  imposes separate
conditions on  $S$ and $T$:
$$\int d^4\!x r^2|T|^2\equiv \int_{-\infi}^{+\infi}dt \int_{0}^{+\infi}d\! r
\; 4\pi r^4 |T|^2\equiv
\int_{-\infi}^{+\infi}dt \int_{0}^{+\infi}d\!r\;
4\pi {L\over r^2}\tt^2 <\infi;$$
$$\int d^4\!x |S|^2\equiv \int_{-\infi}^{+\infi}dt \int_{0}^{+\infi}d\!r
\; 4\pi r^2 |S|^2\equiv
\int_{-\infi}^{+\infi}dt \int_{0}^{+\infi}d\! r\;
  4\pi r^2 {L}^{-3}\st^2 <\infi.$$
Note that
with our choices of $x_i=(-{R\over 2},0,0,0)$ and $x_a=({R\over 2},0,0,0)$,
we have $L(r,t)\neq 0$ almost everywhere on the line $r=0$,
so that a necessary
condition  for normalisation is $\tt (0,t)=0$.

We are thus led  to the boundary value problem:
\bea
 \nabla (h \nabla\tt )=0,&\;\; \hbox{\rm on }\Omega ;\non\\
                 \tt =0,&\;\; \hbox{\rm on }\de\Omega ;\non\\
	         \tt\in H^2(\Omega ),&\;\;\tt\in C^0(\bar{\Omega}),
\label{2_3}\eea
where $\om\equiv\{(r,t)\in {\bf R}^2/r>0\}$ (see Appendix A for definition
of $H^2$).
The operator $\,\nabla^2+\nabla h\nabla \,$ is an elliptic differential
operator
\footnote{A differential linear operator
of order m $P(x,\de )
=\sum_{|\alpha|\leq m}a_\alpha(x) \de^\alpha$
 is called {\bf elliptic}
if
 $\sum_{|\alpha|= m}a_\alpha(x)
p^\alpha \neq 0\;\forall x\in \om \;\; \forall  p\in\rn/\{0\}$;
it is called
 {\bf strongly elliptic}
if
 $\sum_{|\alpha|= m}a_\alpha(x)
p^\alpha>0\;\forall x\in \om \;\; \forall  p\in\rn/\{0\}$.}
 with coefficents analytic in $\om$,
 so that we can apply:

\teo {\it (Elliptic Regularity):
 Let $P(x,\de)$ an elliptic differential operator
in $\om$ with coefficients analytic in $\om$ and let $f$ a function analytic in
 $\om$. If $u\in {\cal D} ' (\om )$ (the space of distributions)
 is a solution of
$ P(x,\de)u=f $ then also $u$ is analytic in $\om$} (see \cite[p.178]{HOR}).

 $\tt $ must thus  be an analytic function in $\om$, hence the following
theorem (see \cite[p.263]{DLA}) applies:

\teo {\it (Strong Maximum Principle): Let $P=a_{ij}\de_i\de_j+b_i\de_i$
be a strongly elliptic operator
with real continuous coefficients on an open connected region
$\om\subseteq\rn$
then:
$\forall u \in C^2(\om )$ such that $Pu=0$ on $\om$ and
 $u$ non constant on $\om$,
$$\inf_\om u <u(x)<\sup_\om u, \;\;\; \forall x\in \om. $$}
In other words,  $\tt$ cannot have global extrema
 on $\om$; using the continuity
on $\bar{\om}$, the boundary conditions Eq.(\ref{2_3}) and the condition
$\tt \rightarrow 0$ when $  r,t \rightarrow \infty, $
it is easy to prove that $\tt =0$.

It follows that $\st $ must be a constant, i.e,
$$ S = {\rm const.} \, L(x)^{-{3\over 2}}. $$
 The normalisation condition however
forces $S =0$. Thus no  left handed normalizable
zero modes exist.   {\it Q.E.D.}

\smallskip
Having  disposed of the zero modes,
 one might then ask if the zero mode of $\dplus $ in the antistanton
field, and that of $\dminus $ in the istanton background, combine in some
way in the valley field so as  to form  a  pair of non zero
 eigenvalue $ \pm \lambda$
of $\dirac$
(a sort of quasi zero mode,  approaching zero
 as the  istanton -  anti-istanton separation
increases). According to the previous result on the essential spectrum,
these (normalisable) modes, if there,  would be embedded in the
continuous spectrum.

Non zero eigenvalues $\pm\lambda$ for $\dirac$ correspond to
  a positive (doubly
degenerate) eigenvalue $\lambda^2$ for the operator $\dirac^2$.
\footnote{It is a well known result   (see \cite{RSA}) that
for a self adjoint operator $T$, $\si_{pp} (T^n) \- =(\si_{pp} (T))^n
\equiv\{\lambda^n/ \lambda\in \si_{pp} (T)$ (where $\si _{pp}$ is the
set of all eigenvalues of $T$).}
The operator $\dirac^2$ is  substantially a Schr\"{o}dinger operator
in $\r4$,
for which the problem of positive eigenvalues is  well studied (\cite{RSB}).
The following result (see \cite{KYOTO} for the general case) is crucial for us:

\teo {\it
Let $P=-D_\mu D_\mu  +c(x)$  ($D_\mu\equiv(\de-igA)_\mu$)
be a differential elliptic operator
in euclidian ${\bf R}^n$ with coefficents analytic almost
everywhere, and such that the field strengths
$F_{\mu\nu}(x)\equiv [D_\mu ,D_\nu ]$ and $c(x)$ go to zero at $x\to \infi$
faster than ${1\over |x|}$, then $P$ cannot have positive eigenvalues.}
\footnote {The theorem
 of Ref.\cite{KYOTO}  refers to  the case of Abelian
backgrounds, but we believe that the  proof is general enough
to be applied to the
non Abelian case, too.  }

Since the operator $\dirac^2$ satisfies all the conditions required,
the presence of   positive eigenvalues can also be excluded.

\section{Large $R/\rho $ }

In the absence of the fermion zero modes,  the anomalous
imaginary part in the forward elastic amplitude cannot be computed in
a straightforward manner as in the instanton calculation of the anomalous
production amplitude (\ref{1_1}), (\ref{1_3}).  Still,
at large $R$ one expects physically that
the generating functional should  reduce to the product
${\cal Z}^{(a)}\cdot {\cal Z}^{(i)}$, where
${\cal Z}^{(a)}$ (${\cal Z}^{(i)}$ ) is the generating
functional in the pure antiinstanton (instanton) background.

Let us introduce complete sets of orthonormal modes $\{ \eta^{(a)}_n \}$
and $\{{\bar \zeta}^{(i)}_n \}$,  $n=0,1,2,....$,
  for the left-handed and right-handed
fermions, respectively.
  They are eigenstates of $ D^{(a)} \dabar$
 and $\dibar D^{(i)} $
\footnote{ The covariant derivatives $D^{(a)},\, D^{(i)}$ are defined
with respect to  the anti\-instanton  $A_{\mu}^{(a)}$ and the
instanton $\,A_{\mu}^{(i)}\, $ of Eq.(\ref{1_6}).  Accordingly
the zero modes are those in the regular gauge (for the lefthanded mode)
and in the singular gauge (for the righthanded one), respectively.
See Appendix D. }:
$$  \dabar \etm = {\bar k_m} \zema\,\, (m=0,1,...),
  \qquad  D^{(a)} \zema = k_m \etm\,\,  (m=1,2,...),       $$
\beq {\bar D}^{(i)} \emi =  l_m \zetm\,\, (m=1,2,...),
  \qquad
D^{(i)} \zetm = {\bar l_m} \emi\,\,  (m=0,1,...), \label{3_1}   \eeq
where
\beq {\bar k_0}={\bar l_0}=0. \label{3_2}\eeq
The functional integral can  then  be defined as:
\footnote{In Ref.\cite{15} the valley approach has been applied also to
 fermions, to perform the functional integration.
The authors also use  a projector to select the anomalous
intermediate states, which however is not explicity
 defined. It is not clear to us how
such calculations can be done in practice, but we expect nontrivial corrections
at finite $R$ to the leading term (\ref{3_15}).}

\beq\int{\cal D}\psi{\cal D}{\bar \psi} \equiv \prod_{m,n=0} da_m\,
d{\bar b}_n;  \label{3_3}\eeq

\beq\psi(x)  = \sum_{m=0}^{\infty} a_m \eta^{(a)}_m(x),\qquad
{\bar \psi}(x) = \sum_{n=0}^{\infty} {\bar b}_n {\bar \zeta}^{(i)*}_n(x).
\label{3_4}\eeq

As is clear from the way Eqs.(\ref{3_1})-(\ref{3_4}) are written,
we first  put  the system in a large but finite  box
of linear size $L$
( such that $ L \gg  R, \rho $)
so that all modes are discrete.  After the derivation of  Eq.(\ref{3_8}) below
(i.e., after the sum over the complete sets is done),  however,
$L$  can be sent to infinity without any difficulty.

The two point function $ I(x,y)  $ can be written as
\bea
 I(x,y) &=& \det{\bar D}\, \langle x|\dinv|y \rangle\non\\
      &=&\det {\bar D}\, \{ \langle x|a,0 \rangle
\langle a,0|\dinv |i,0\rangle \langle i,0|y \rangle
    +\sum_{m\ne 0}\langle x|a,m\rangle
\langle a,m|\dinv|i,0\rangle \langle i,0|y \rangle\non\\
     &+&\sum_{n\ne 0}\langle x|a,0 \rangle
\langle a,0|\dinv|i,n\rangle \langle i,n|y \rangle
      +\sum_{m,n\ne 0}\langle x|a,m\rangle
\langle a,m|\dinv|i,n\rangle \langle i,n|y \rangle \}:\non\\
&&
\label{3_5}
\eea
the term proportional to the product of the zero modes
 has been singled out.   We wish to compute Eq.(\ref{3_5}) at small
$\rho/R$. To do this, first
let us write the operator $\dbar$ in the above basis:
\beq \dbar = \pmatrix{d&v_1&\ldots&v_n&\ldots\cr
                         w_1&X_{11}&\ldots&X_{1n}&\ldots\cr
                 \vdots&\vdots&\ddots&\vdots&\ddots\cr
                   w_m&X_{m1}&\ldots&X_{mn}&\ldots\cr
                  \vdots&\vdots&\ddots&\vdots&\ddots. \cr} \label{3_6}\eeq
where we defined
\beas
 \,d &\equiv& (\dbar)_{00}= \bra i,0|\dbar |a,0 \ket =\bra i,0|\cbar|a,0 \ket ,
\\
 v_n &\equiv&  (\dbar)_{0n}= \bra i,0|\dbar |a,n \ket =
\bra i,0|\bbar|a,n \ket ,\\
  w_m &\equiv& \bra i,m|\dbar |a,0 \ket =\bra i,m|\cbar|a,0 \ket ,\\
X_{mn}&\equiv&\bra i,m|\dbar|a,n \ket .
\eeas
Operators   $\cbar \equiv C_{\mu} {\bar \sigma}_{\mu};\,\,
\bbar \equiv B_{\mu} {\bar \sigma}_{\mu}$  are defined by:
\beq D_{\mu}^{(valley)} = D_{\mu}^{(a)} + C_{\mu} = D_{\mu}^{(i)} + B_{\mu}.
\label{3_9} \eeq
The explicit expression of these quantities are found in Appendix D.

The idea is that the matrix elements involving either of the zero modes,
$d,\, v_n,\, w_m, $ are all small by some overlap integrals (see
Footnote 13) while the
matrix elements $X$ are large because the wave functions of non zero modes
are extended to  all over the spacetime.

The inverse matrix $\dinv$ is given by:
\bea
 (\dinv)_{00} &=& 1/( d - vX^{-1}w ) \non\\
 & = & d^{-1} + d^{-2}v_m (X^{-1})_{mn} w_n + \cdots;\non \\
(\dinv)_{mn} &= &(X - {1\over d}w \otimes v) ^{-1}=
X^{-1} ( 1 - {1\over d}w \otimes v X^{-1} )^{-1} \non\\
 &= &(X^{-1})_{mn} + d^{-1}(X^{-1})_{ml} w_l v_k (X^{-1})_{kn}
        + \cdots,\non\\
 (\dinv)_{0n} &=& - d^{-1} v_l (\dinv)_{ln},\non \\
(\dinv)_{m0} &=& - (\dinv)_{00} X^{-1}_{mk} w_k,  \label{3_7}
\eea
where $X^{-1}$ is the inverse of the submatrix $X$ in the space orthogonal
to the zero modes.

Inserting Eq.(\ref{3_7}) into Eq.(\ref{3_5})
 and after some algebra (see Appendix B   for
derivation) one finds a remarkably simple (and still exact)
 expression for $I(x,y)$:
\bea
 I(x,y) &= \det X\,\{\bra x|a,0\ket -
 \bra x | X^{-1}\cbar|a,0\ket \}\, \{\bra i,0|y\ket -
\bra i,0|\bbar X^{-1} |y \ket \}  \non\\
 & + \det \dbar \, \bra x | X^{-1} | y \ket . \label{3_8} \eea

Unfortunately, the inverse $X^{-1}$ is not known, and our present knowledge
does not allow us  to compute it perturbatively in any small parameter.

In Ref.\cite{32}, we used a perturbative formula for $X^{-1}$
in terms of the propagator in the anti-instanton background $\sbar$,
 to get
an expansion for $I(x,y)$.
 A closer look
at various  terms arising from such an expansion, however, has revealed
several  difficulties with this procedure. First, there are difficulties in
the application of  the  LSZ procedure on some terms,
caused by the infrared behaviour  of $\sbar$.  Also, the expansion
resulting from  Eq.(12) of \cite{32} (obtained from Eq.(\ref{3_8}) by use of
the
above mentioned formula for $X^{-1}$ )
 turns out not to be an expansion in $\rho/R$ contrary to
the incorrect claim made there.

Nonetheless, we believe that Eq.(\ref{3_8}) displays  the main features of the
two point function in the valley background correctly.
  The effect due to the zero
modes is separated explicitly and everything else is expressed by the
smoother two point function,
\beq S^{\prime}_{x,y} = \bra x| X^{-1} | y \ket. \label{x}\eeq

In order to find the amplitude, one must apply the LSZ reduction on
the four point function $ I(x,y) I(u,v). $  This requires only the knowledge
of $\sprime$ at large $x$ and $y$.

We assume $\sprime$ to behave
 at large $x$ and $y$ (with $x_i$ and $x_a$ fixed)
 as
\bea
 \sprime & \sim & U^{\dagger}(x) S_F(x,y)  U(y),\label{ipotesi}\\
   U(x) & =& { {\bar \sigma}_{\mu} (x-x_a)_{\mu}
\over \sqrt{(x-x_a)^2}}, \non
\eea
 where $S_F$ is the free Feynman propagator. This behaviour is suggested by the
 fact that the valley field has a pure gauge form at large x,
\beq   A_{\mu}^{(valley)} \sim {i\over g}\, U^{\dagger}\partial_{\mu}\,U
   \sim O({1\over x}).   \label{3_10} \eeq

To proceed further we assume that
\beq \det X / \det {\bar \partial}  = {\rm const.}    \label{3_10bis}\eeq

\noindent
Next the ratio  $\det \dbar / \det X $ can be estimated as follows
(see (\ref{3_7})):
\beq {\det \dbar \over  \det X} = ((\dinv)_{00})^{-1} \simeq d
\simeq {\rm const.}\, \rho^2/R^3,   \label{3_10ter}\eeq
where use was made of
$$ d =  \dbar_{00}= \bra i,0|\cbar| a,0 \ket =
 \int_z \zet0\!(z)^*  \cbar\!(z)\, \et0\!(z)
   \sim \rho^2/R^3. $$

\noindent
Combining Eq.(\ref{3_10bis}) and Eq.(\ref{3_10ter}) gives
\beq
{\det \dbar \over  \det {\bar \partial}} \sim  \rho^2/R^3. \label{3_10quater}
\eeq

 With Eq.(\ref{3_10quater})
and Eq.(\ref{ipotesi}) in Eq.(\ref{3_8}) one can extimate the amplitude
and the leading contribution to its anomalous part.
  A naive application of the standard LSZ procedure, i.e.,
\beq \lim_{ q^2 \to 0}
   \int d^4\!x \,e^{iq\cdot x}\, {\bar u}(q) \,{\bar \sigma}_{\mu}
\partial_{\mu}\, \bra T \psi(x) \cdots \ket \cdot  \label{3_11} \eeq
however  leads to a number of difficulties.  On the one hand, a
large correction is  found from the second term of
$\bra i,0|({\bf 1} -
\bbar X^{-1})| y \ket $.   The anti-instanton zero mode
 (in the regular gauge)
$\bra x| a,0 \ket $,
  on the other hand, has not the right asymptotic form of the free
propagator.

All  the problems however can be attributed
to the inappropriate choice of the gauge, Eq.(\ref{1_6}).
Indeed, we have already noticed that  $A_{\mu}^{(valley)}$
behaves asymptotically as a pure gauge field, Eq.(\ref{3_10}).
It is then not surprising that the naive LSZ fails: the fermions never
become free, whatever distance they travel, a situation somewhat
reminiscent of the Coulomb scattering.

  Note that  if the fermions were moving
in a single instanton field, it would have
been sufficient to work in the singular gauge, which also satisfy the
Lorentz gauge condition.   For the valley field instead,
 a gauge transformation  to a sort of double singular gauge
\bea
  {\tilde A}_{\mu}^{(valley)} &  = &
U ( A_{\mu}^{(valley)} + { i\over g}
\partial_{\mu} ) U^{\dagger},\non \\
   U & =& { {\bar \sigma}_{\mu} (x-x_a)_{\mu}
\over \sqrt{(x-x_a)^2}},  \label{3_12}
\eea
 in which both instanton and
anti\-instanton do drop off faster at infinity,
takes us out of the Lorentz gauge.

  We shall by-pass this problem  by computing the Green function
in the original gauge Eq.(\ref{1_6}) and by transforming it to the more
physical
gauge Eq.(\ref{3_12}) only at the end of calculation,
  before applying the LSZ reduction.
Note that the two point function $I(x,y)=\bra T \psi(x) {\bar \psi(y)}\ket $
 transforms covariantly,
$$ I(x,y) \longrightarrow  U(x) I(x,y) U^{\dagger}(y).  $$
 The correct  LSZ procedure is thus:
\bea
&& \lim_{ q^2 \to 0} \int_y  \bra  \cdots | y \ket\,
 U^{\dagger}\, {\bar \partial}\, u(q)\, e^{-iq \cdot y}    \non\\
&=& \lim_{ q^2 \to 0} \int_y   \bra  \cdots | y \ket\,
 \{{\bar \partial} - ( {\bar \partial}
U^{\dagger} ) U \}\, U^{\dagger}\, u(q)\, e^{-iq \cdot y}, \label{3_13}
 \eea
for the initial fermions;
\bea
& & \lim_{p^2 \to 0} \int_x  e^{ip \cdot x}\, {\bar u}(p)\,
 {\bar \partial}\,
U \, \bra x |\cdots \ket \non \\
  &=& \lim_{p^2 \to 0} \int_x  e^{ip \cdot x}\, {\bar u}(p)\,
 U \,\{ {\bar \partial} +
  U^{\dagger} ({\bar \partial} U) \}\, \bra x| \cdots \ket,
 \label{3_14}
\eea
for the final fermions.

We analyse  first the contribution
\beq \label{fattorizzato}
\{\bra x|a,0\ket -
 \bra x | X^{-1}\cbar|a,0\ket \}\, \{\bra i,0|y\ket -
\bra i,0|\bbar X^{-1} |y \ket \} \eeq
to the two point function (\ref{3_8}).

Considering the term $\{\bra x|a,0\ket -
 \bra x | X^{-1}\cbar|a,0\ket \}$  and applying the LSZ amputation,
we obtain:
\beq
{\bar \partial} U(x)\, \et0\!(x)- U(x){\bar C}(x) \et0\!(x)
\label{left},\eeq
where we used (\ref{ipotesi}) and
${\bar \partial}(S_F)_{x,y}  = \delta(x-y)$.
Note that the righthanded zero mode $\et0\!(x)$
  is automatically transformed into the singular gauge form
$U(x)\, \et0\!(x)$ which has the correct asymptotic behaviour.

Also, the  large correction  arising
 from the second term of
$\bra i,0|({\bf 1} - \bbar X^{-1})| y \ket, $
due to the bad asymptotic behaviour of $B_{\mu} = D^{(valley)}_{\mu}
 - D^{(i)}_{\mu}$
gets automatically cancelled.
 To see how this occurs, first write
$$ B_{\mu} =-ig a_{\mu} + B_{\mu}^{\prime}, $$
$$  a_{\mu}\equiv {i\over g}U^{\dagger}\,\partial_{\mu}U ;\,\,\,
          B_{\mu}^{\prime}\equiv -ig (A_{\mu}^{(a)} -
 a_{\mu} + A_{\mu}^{(int.)}).$$
Then
$$B_{\mu}^{\prime} = O(1/x^2), $$
asymptotically.  Then
\bea
&  & [\bra i,0|y\ket - \bra i,0|\bbar X^{-1} |y\ket] U^{\dagger}
{\bar \partial}\non \\
   & \simeq &\bra i,0|y\ket {\bar \partial} U^{\dagger} +
\bra i,0|y \ket  (U^{\dagger} {\bar \partial} U) U^{\dagger} \non\\
 & -& \bra i,0|y \ket [ (U^{\dagger} {\bar \partial} U) + {\bar B}^{\prime}]
 U^{\dagger} S_F {\bar \partial}  \non\\
 &=&
\zet0\!(y)^* {\bar \partial} U^{\dagger}(y) + \zet0\!(y)^*
{\bar B}^{\prime}(y)
U^{\dagger}(y),\label{right}
\eea
where use was made of the identity, $(S_F)_{x,y} {\bar \partial}_y =  \delta
(x-y).$

Multiplying (\ref{left}) and (\ref{right}) and doing the Fourier transform
prescribed by LSZ we obtain the contribution of the first term of
(\ref{3_8}) to the amplitude.
The leading term in this amplitude, containing
$ \rho^2  \exp(ip \cdot x_a ) \exp(-iq \cdot x_i) $, arises from the
product of the
zero modes in Eq.(\ref{fattorizzato}), as seen from the asymptotic
 behaviour of the latter,
$$
\et0(x)\sim {\rho \over (x-x_a)^3} ;\quad \zet0(x) \sim
{\rho \over (x-x_i)^3}. $$
This contribution to the amplitude must obviously be considered as anomalous.
All remaining terms in (\ref{fattorizzato})
give a contribution of order $O(\rho/R)$
\footnote{Clearly suppression is due to the appearence of product of functions
with small overlapping support or to the explicit factor ${\rho \over R}$
in $A_\mu^{(int.)}$ (see appendix D for its explicit form).}
or less
to the total amplitude and so are negligible for
large $R$.
In particular, second term of Eq.(\ref{right})
 gives a $O(\rho/R)$ contribution which is still
proportional to $ \rho^2  \exp(ip \cdot x_a) \exp(-iq \cdot x_i) $:
such a correction  to the anomalous part of the elastic amplitude can become
important  when $R\sim \rho$.

The second term of (\ref{3_8}),
with the assumption (\ref{ipotesi}) for $\sprime$, on the other hand,
 gives clearly a non anomalous
contribution to the amplitude. Also eventual subleading
asymptotic terms of $\sprime$
contributing to anomalous part would be suppressed by the small factor
(\ref{3_10ter}) compared to the leading anomalous term previously found.

\smallskip
Recapitulating, the  leading anomalous contribution ( for $R/\rho \gg 1$)
is essentially given by
 \beq I^{(anom)}(x,y)
 \simeq   {\rm const.}\, \et0\!(x)\,\zet0\!(y)^*.  \label{3_15} \eeq
Inserted in the four point function Eq.(\ref{1_9}), it leads
  to the "anomalous part of the forward
elastic amplitude", required by unitarity and chiral anomaly.

To reach the above conclusion really, however, we must make
 one further check. For
the first term of Eq.(\ref{3_8}) to represent the anomalous process,
the intermediate state must contain fermions satisfying the instanton
selection rule,
Eq.(\ref{1_3}).

That this is indeed so can be seen from  Eq.(\ref{3_10quater}). The functional
integration
yields, for each flavour ($i=3,4,\cdots
 N_F$) one factor of
$$\det \dbar / \det {\bar \partial}  \sim \rho^2/R^3, $$
but this is precisely the factor expected for a left-handed fermion,
 produced at the instanton center (with amplitude
$ \rho $ ), propagating backward to the antiinstanton  (with amplitude, $
\sim 1/(x_a-x_i)^3 = 1/R^3 $ ) and absorbed by the latter (with amplitude,
 $\, \rho $ ), see \mbox{Fig. 1}.
(This provides an a posteriori justification of the assumption
Eq.(\ref{3_10bis}) for $\det X$.)

To be even more explicit, suppose that we are going to observe the final state
by setting up an appropriate detector. This would correspond to introducing
a source (or sink) term for each flavour,
$\int dx\,J^i_{\mu}(x)\psi_i(x) + h.c., $
and taking the first derivative with respect to the sources. This would
produce  pairs of zero modes (as in Eq.(\ref{3_15})):
 the fermions required by the
topological selection rule are indeed there, as long as $R/\rho$ is large
(see \mbox{Fig. 2}).

In this connection, it is amusing to note  how a sort of dilemma between the
factorisation in the flavour  of the functional integration Eq.(\ref{1_9})
and the topological selection rule, is solved in the semi-classical
approximation we are working with. Consider as an illustration  the forward
elastic amplitude with two initial particles of the same flavour,
$1 + 1 \rightarrow 1 + 1.$   In this case  no anomalous
piece is expected to be present. To see this, let us note first that
  due to the Grassmanian
nature of the fermion fields and sources one gets the amplitude with the
Fermi-Dirac statistics automatically  built in.
Due to its factorized form, contribution of Eq.(\ref{3_15}) to the four
point function  cancels out
  by  antisymmetrisation in $x \leftrightarrow
u;\, y \leftrightarrow v$.

\smallskip
As a further check, consider the case of massive
fermions (we have so far regarded them massless). For definiteness,
consider Dirac type masses: we write
$ \psi_1 = \psi^{(1)}_L; \psi_2 = \psi^{(1)c}_R;
 \psi_3 = \psi^{(2)}_L; \psi_4 = \psi^{(2)c}_R;$ etc. The contribution to
${\cal Z}$ of a single Dirac flavour is given by:
$$ {\cal Z} = \det  \pmatrix{m{\bf 1} & i D  \cr
             i \dbar & m^* {\bf 1} \cr}     $$
where the matrix acts
 between  $ {\bar \psi}=({\bar \psi}_1, \psi_2)$
 and $\psi =({\bar \psi}_2, \psi_1)$.

At large $R/\rho$ we know that the first column and the first row of
$\dbar$ and $D$, regarded as matrices in our  basis (Eq.(\ref{3_1})),
 become small due to the tiny overlap integrals, while the
rest of the matrices remains large. It then follows that
\beq \lim_{R/\rho \rightarrow \infty} {\cal Z}
 = {\rm const.}\, m\, m^*,
  \label{3_16}\eeq
which is  consistent with the clustering for massive but
sourceless fermions.  This argument further implies that
Eq.(\ref{3_16}) is a good approximation when
\beq {\rho^2 \over R^3 } \ll |m| .    \label{3_17}  \eeq

 \section{Small $R/\rho$. }
\subsection{Instanton -Anti-instanton Melting. }

The approximation (\ref{3_15})  fails at small $R/\rho \le 1 $ for
obvious reasons.

In particular, in the limit  $R/\rho \rightarrow 0 $,
the classical field $A_{\mu}^{(valley)} $ reduces to the trivial,
perturbative vacuum. In the free theory, the procedure adopted above is
still formally valid, but the appearance of the product of the zero modes
(as in Eq.(\ref{3_5})) is of course a fake, the total two-point function being
simply $\sigma_{\mu} (x-y)_{\mu}/(x-y)^4.$  This means that at
certain $R/\rho \sim O(1) $ the leading anomalous term Eq.(\ref{3_15}),
 must be effectively cancelled by contributions from  other
terms. Altough we do not know exactly at which value of $R/\rho$ this occurs,
there is a fairly
good indication that such a transition takes place around $R/\rho\sim 1$.

We have in fact computed numerically
the integral up to
 the Euclidean time $x_4$ of the topological
density,
\beq
 C(x_4) = -\int_{-\infty}^{x_4} \int d^3x {g^2\over 16 \pi^2}
\hbox{\rm Tr }F_{\mu \nu}
 {\tilde F}_{\mu \nu} = {\cal N}_{CS}(x_4)-{\cal N}_{CS}(-\infty),
 \label{4_1} \eeq
as a function of $x_4$ for several values of $R/\rho$,
for the valley background of Eq.(\ref{1_6}). (See Fig.3.)
    ${\cal N}_{CS}(x_4)$ is
 the Chern Simons number
\beq {\cal N}_{CS}(x_4)\equiv -\int d^3 \!x {g^2\over 16 \pi^2}
\epsilon^{4ijk}\hbox{\rm Tr }(F_{ij}A_k-{2\over 3} A_i A_j A_k ).\eeq
  The instanton and antiinstanton are
situated at $ ({\bf 0}, R/2) $ and at $ ({\bf 0}, -R/2) $, respectively.

In Fig.4  we plot also the behaviour of the maximum of each curve,
corresponding to $C(0)$, as a function of $R/\rho$.  The behaviour of $C(0)$ is
powerlike both
at large and small R:
\beas
C(0) &\sim& 1 - 12(\rho/R)^4, \quad R/\rho \gg 1, \\
C(0) &\sim&  {3\over 4} (R/\rho)^2 ,\quad R/\rho \ll 1 .\eeas

Actually,  the exact expression for $C(x_4)$ in terms
of $x_4$ can be found by using conformal transformations,
\cite{prov}. In particular it can
be proved that
\beq
C(0)=3({z-1\over z+1})^2-2({z-1\over z+1})^3\eeq
where $z$ is defined in (\ref{1_7}).

It can be seen from Fig.3   that the topological structure
is well separated and localised at the two instanton centers
only at relatively large values of $R/\rho$,   $R/\rho \ge 10 $.
At very large $R$,   $C(x_4)$
approaches  the product of two theta functions.
Vice versa, for
small $R/\rho \le 1 $  the gauge field is seen to
collapse to some insignificant fluctuation
around zero, not clearly distinguishable from ordinary perturbative ones.
In this case there will be no level-crossing \cite{34,35,crossing} hence
 no chiral anomaly.

Fig.3 and Fig.4 indicate that the instanton and anti\-instanton start
to melt at around $R/\rho \simeq 5 $  and go through the transition
quickly, the center of the transition being at
 around $R/\rho =1$ .

The behaviour of $C(x_4)$ at small $R$ suggests that at $R/\rho \le 1$
one can compute $I(x,y)$  by simply using the standard perturbation
 theory, with
$A_{\mu}^{(valley)}$ as perturbation.
Our
 results have several interesting implications.

\subsection{
Fermion Number Violation
in the TeV Region Scattering Processes.}

According to the method developed in Refs.\cite{6,8,9,11,15,22,25},
 taking into account the
contribution of the Higgs field in the action and using the saddle point
approximation in the integrations over the collective coordinates, one finds
a relation among the c.m. energy $\sqrt s $ and the saddle point
values of the parameters $R$ and $\rho$.
It was found in Ref.\cite{15}, in a simple toy-model calculation
 which uses the valley trajectory of Eq.(\ref{1_6}),
 that at an energy of the order of the sphaleron
mass the total action  (at the saddle point) vanishes
and that the 't Hooft suppression factor disappears.

But at that point,  the corresponding saddle point values of the
instanton parameters are found to be   $ \rho = 0;\, R/\rho = 0 $ (hence
$A_{\mu} = 0 $) : the unsuppressed cross section should simply correspond
to a non-anomalous, perturbative cross section. This was pointed out also by
some authors \cite{17}.

What happens is that the anomalous term of Eq.(\ref{3_15})
disappears at some $R/\rho$. Our numerical analysis
of the Chern Simons number shown above suggests that this
occurs quickly at around $R/\rho=1$.   And this turns out to be
precisely where the valley action sharply
drops to zero: see Fig.5 taken from Ref.\cite{15}.  We must conclude that
 the toy-model considered there  shows no
reliable sign of  anomalous cross section (associated with the
production of large number ($\sim 1/\alpha $) of gauge and Higgs bosons)
becoming large.

Of course, there is no proof that all sorts of quantum correct\-ions to the
inst\-anton-induced process (\ref{1_1}) are
 effectively described by a  classical
background such as Eq.(\ref{1_6}).   But this is another issue.
The point here is that there are as yet no calculations anyway
which show that the anomalous process becomes observable in high energy
scatterings.

More generally, the consideration of this paper suggests that, in order for
fermion number violating cross sections to become observable at high
energies, a new mechanism must be found in which the
background field governing the elastic amplitude Eq.(\ref{1_2})
does not effectively reduce to $ A^{eff}_{\mu} = 0 $.
 Note that this is necessary,
whether or not multiple-instanton type configurations become
important.   It is  difficult to envisage
 such a novel mechanism, not accompanied by some finite fraction of the
't Hooft factor.

\subsection{Theories without Fermions.}
It is often stated that fermions are not essential,  as dynamical
effects of inst\-anton-induced cross sections can well be studied in a theory
without fermions.
This is certainly true, but this does not mean
that the consideration of this paper is
irrelevant in such a case.

Quite the contrary.  The crucial factor,
$$\exp(ip_j\cdot x_a) \exp(-iq_j\cdot x_i) \qquad (j=1,2) $$
associated with the external particles, in the case of process with fermions,
appears upon LSZ amputation applied to the product
of the zero modes of Eq.(\ref{3_15}).  In the case of external gauge bosons,
the same factor emerges as a result of the "semiclassical" approximation,
\bea
& <TA_{\mu}(x) A_{\nu}(u) A_{\rho}(y) A_{\sigma}(v)> \non\\
& \simeq A^{(a)}_{\mu}(x) A^{(a)}_{\nu}(u) A^{(i)}_{\rho}(y)
 A^{(i)}_{\sigma}(v)  + .... \label{4_2}
\eea

It is however clear  that this approximation, which is probably good at
large $R/\rho$,  should fail at  $R/\rho < 1 $  just as Eq.(\ref{3_15}) does.
The first term of Eq.(\ref{4_2}), the anomalous term,  disappears
 precisely when the
cross section is claimed to become observable.

Thus the use of the theories without fermions  does not change our
conclusion: there is no evidence for the anomalous process to become
observable in the TeV region scattering.
   The scarcity of our knowledge is actually even worse. In fact,
 in the existing literature the substitution Eq.(\ref{4_2})
is just made
{\it by hand }: no proof of Eq.(\ref{4_2}) seems to be known. It is an
interesting question
how our approach (of Section 3) can be generalised to the case of external
bosons.
\bigskip

\section { An Apparent Paradox and its Resolution.}
It is quite remarkable that the fermion number violating
term (Eq.(\ref{3_15})) at large instanton-antiinstanton separation,
 emerges without
there being a single, dominant mode of the Dirac operators, $\dbar$, or
 $D$.
The result Eq.(\ref{3_10quater}) for $\det \dbar$  neither implies the
existence of a
particular eigenmode with eigenvalue, $\sim \rho^2/ R^3 $,  nor requires
that such
a mode dominate over others.

Indeed, for any finite $R$ we have  established in Section 2 the following.
First of all,
neither  exact zero mode  nor quasi-zero modes
exist in the valley background.
Secondly, there are many non-zero modes, definitely lying below
  $\rho^2/ R^3$, and forming a continuous spectrum,
reaching down to $0$.   In
particular,
putting the system in a large box of fixed size $L$ such that
\beq L\gg R, \rho \label{4_3},\eeq
for the lowest lying  modes with  $kR \ll  1 $
 both
the eigenvalue and the wave function differ little  from the free
spectrum, as shown perturbatively in Appendix C.

(In passing, this shows that a single $i-a$ pair in itself cannot lead to
chiral symmetry breaking, $<{\bar \psi} \psi > \ne 0, $ in QCD: the latter
necessarily  requires
an accumulation of  eigenvalues towards $0$ (see Ref. \cite{36}).
 In the context of
instanton physics, that would require something like
the  "instanton liquid" (Ref.\cite{37}).)

The results such as Eq.(\ref{3_15}) and Eq.(\ref{3_16}) are thus  collective
effects in which many modes contribute together;  no single mode
of $\dbar$ or $D$ plays any
particular role.
 We then seem to face a somewhat paradoxical situation. At large instanton
- antiinstanton separation, physics must factorise  and
we do find results
consistent with such intuition. This is fine. The problem is that
mathematics to achieve this looks
very different from that of the usual instanton physics where a single
fermion zero mode plays a special role. In our case, there is no hint even of
the presence of a quasi zero mode.
What is going on?

The key to the resolution of this apparent paradox is the
inequality, (\ref{4_3}). In order to be able to compute the S-matrix
elements from the four point function (LSZ procedure),
we are in fact forced to work in a spacetime region whose linear size
 is much larger than
the physical parameters $R$ and $\rho$, independently of the ratio $R/\rho$.
  All
fields must be normalised in such a box and functional integration defined
there. This makes our system with instanton-antiinstanton  background
always distinct and not continuously connected to the system with a single
(say)  antiinstanton. The gauge field topology
remains firmly in the trivial sector.

In spite of our use of the "zero modes" $\et0\!,\, \zet0$  as a convenient
device for calculation,  they are  not
a good approximate wave function for the lowest modes in the valley
background, however large $R/\rho$ may be.
In fact, if one insisted upon using $\et0$ as the "unperturbed" state,
one would discover that the effect of $\cbar$ is always
non\-perturbative and large (near $x_i$). \footnote{ We did verify that
the standard
perturbation
theory applied to  $H = D \dbar = D^{(a)} \dabar \,  + H^{\prime}$
with $H^{\prime}=D^{(a)}\cbar + C \dabar + C\cbar$,
yields  $\Delta E =0 $ to all orders,
 reflecting the topological stability of the
fermion zero mode. (Another way to see this is to notice a
supersymmetric structure underlying the system. We thank C.Imbimbo for
pointing this out to us.)
Such a result, however, is false in the case of topol\-ogy-
changing modification of the gauge field, in view of the result of Section 2. }

All this is clearly to be distinguished from the situation where we ask
e.g., what the effects of an instanton on a distant planet are.
\footnote{A somewhat related consideration
is found in Ref. \cite{36}.}  Such a case would
correspond to the inequality opposite to (\ref{4_3}),
$$ R \gg L, $$
if we restrict ourselves to the instanton-antiinstanton case.  As long as
we are interested in physics inside our laboratory (or on the earth, anyway)
both physics and mathematics are
described by just ignoring the distant instanton,
to a good approximation.  (Particles are produced and detected in the
region of volume, $L^4 \ll R^4$,
 the gauge field having the winding number $-1$,
fields and functional integrations defined in the same "box", etc.)
The effect of the distant instanton is a true, and negligibly small,
perturbation in this case.

\section{ Conclusion.}
In this paper we study the fermions propagating in the background of
instanton-anti-instanton valley.  The spectrum of the Dirac operator is
studied first, proving the absence of the zero mode for all values of the
collective coordinates.  We  then study the
fermion Green function at large  $R/\rho$ (the instanton and anti-instanton
well separated).
This solves the problem raised in the introduction of unitarity versus
chiral anomaly, yielding the leading  anomalous part of the
forward elastic amplitude.  For small $R$ our study remains somewhat
indirect, being limited for the moment to the numerical  study of Chern
Simons number as a function of the time and of the ratio $R/\rho$.
Nonetheless the latter has several interesting implications as was
discussed in Section 3.

Applied to the question of fermion number violation in the TeV region
scattering in the standard electroweak theory, our results imply that
there is as yet no theoretical evidence that such a process becomes
observable, contrary to the claim of several papers.

In the application to Quantum Chromodymamics,
we hope that our study serves as the starting point of an improved
 study of the instanton liquid model for the QCD vacuum.

\bigskip
\noindent {\bf Acknowledgment.} The authors are grateful to D.
Amati, M. Bertero, C. Imbimbo, H. Leutwyler, G. Morchio, M. Maggiore and
V. Zakharov
for interesting discussions.

\appendix
\section { Essential Spectrum
 of the Dirac Operator in the Valley   }
In this Appendix we choose for the valley field the 'clever gauge'
introduced in Section 2:
$$A_\mu =-{i\over g}\sb_\rho (\si _\mu \sb _\nu -\delta_{\mu \nu})\si _\tau
{ v_\rho v_\tau \over v^2}\,
 H_\nu $$
$$ H_\nu =
 {x-x_a\over (x-x_a)^2+\rho ^2}-{x-x_i\over (x-x_i)^2+\rho ^2}
\;\;\;\;\;\;( v=x-x_i+y )$$
which is obtained  from Eq.(\ref{1_6}) by the gauge transformation,
$U={\sb _\mu v_\mu \over \sqrt{v^2} } $.
 $A_{\mu}$ is a bounded function and
$A_\mu (x)\sim {R_\mu\over |x|^2}$ as $x \to \infi$, (which implies
the uniformity in direction of $x$ of convergency at infinity).
 The following theorem can
then  be used:

\teo{\it If $A_\mu(x)$ is a bounded function going to $0$ at $\infi$,
uniformly
in $x$, then the essential spectrum
of $\dirac =i \gamma_\mu (\partial_\mu -ig A_\mu) $ in euclidian
space $\r4$ is
$\si _{ess}= (-\infty ,\infty) $.}

\smallskip
\noindent {\bf Proof:}
\smallskip
We  treat the vector potential term as a perturbation of
the free Dirac operator, and we prove that complete and the free Dirac operator
have the same essential spectrum.
It is a well known result \cite[cap. IX]{DLB} that the free Dirac operator
$$\dirac _0\equiv i \gamma _\mu \de_\mu :L^2 (\r4 )^4 \rightarrow  L^2
(\r4 )^4$$
(that is acting in the space of square-integrable four-spinors),
with domain $D(\dirac _0 )= H^1 (\r4 )$
\footnote{We indicate with $H^m(\rn )$ the  space (called
Sobolev space, see \cite{DLA}) of all functions
which are in $L^2(\rn )$ together with their first m (distributional)
derivatives. These spaces, if endowed with an adequate topology are Hilbert
spaces.}
is self-adjoint (and so closed) with spectrum
$$\si (\dirac _{0} )\equiv \si _{ess}(\dirac _{0} )\equiv (-\infi ,+\infi)$$
Thanks to the  bounded\-ness of vector potential the operator
$A:L^2(\r4)^4 \rightarrow  L^2(\r4)^4$
which acts multiply\-ing spinors by $g\gamma_\mu A_\mu(x)$, is bounded
\footnote{That is $\nrm A \psi\nrm < B\nrm \psi \nrm \forall \psi$,
property which is equivalent to continuity.}
and
self-adjoint, so that the
operator $\dirac\equiv \dirac_0 + A$
is self-adjoint (with $D(\dirac )=D(\dirac _0)$).

To prove the equality of the essential spectra we use the following
theorem (see \cite[ p.114 and p.116]{RSB})
with $T\equiv \dirac_0$ and $V\equiv A$:

\smallskip
\teo {\it Let $T$ be a self-adjoint operator in a Hilbert space, and $V$ a
 bounded  self-adjoint operator. Also let $V$ $T^m$-compact for some integer
$m>0$.
Then
$$ \si_{ess}(T+V)=\si_{ess}(T) $$}
An operator  $V$ is called  compact relatively
to a self-adjont operator $T$
(shortly  {\bf T-compact})
 if $D(T)\subset D(V)$  and the operator
$V(T-\mu)^{-1}$ is compact, for some $\mu\not\in \si (T)$.
We shall not give  the definition of {\bf compact} operator here
(see \cite{RSA}),
but  just use the result   that in $\rn $
an integral operator with kernel $K(x,y)$  such that
\beq \int\int d^n \!x \;d^n \!y |K(x,y)|^2<\infi \label{D_1}\eeq
 is compact.

 In order to use this theorem with $T\equiv\dirac_0$,
 $V\equiv A$, $m=4$ and $\mu =-1$  we must prove that
 the operator
$A (\de^4+1)^{-1}$
 is compact.
\footnote{Note that, we use $T=\dirac_0$, therefore
$T^4\equiv \de^4$. The following proof is inspired to \cite[p.117]{RSB}. }
To do this consider the sequence of truncated vectors potentials
$${A_n}_{\mu}(x)\equiv A_\mu (x) \;\theta(n-|x|), \;\;\;n=1,2,...$$
which, due to the property of (uniform) convergence to zero
at infinity  of the vector
potential,
 converges to $A_\mu (x)$ uniformly
 in all $x$.
\footnote{I.e. in the $\nrm \;\nrm _{\infi}$ norm.}
 This guarantees that
also the sequence of associated multiplicative operators
$A_n\equiv g\gamma_\mu{A_n}_{\mu}(x)$ converges
to $A$ in the (bounded) operators
space,
\footnote{For operators that acts multipling by a function,
the norm in the operator space  coincides with the
$\nrm \;\nrm _{\infi}$ norm of that function.}
 and, by continuity of multiplication,
 the sequence  $A_n(\de^4+1)^{-1}$  converges to
$A(\de^4+1)^{-1}$ also.

 Because the space of compact operators is a closed space,
it then suffices to prove that for all $n$ the
 operators $A_n(\de^4+1)^{-1}$ are compact.
But these operators are integral operators with kernel:
$$K_n(x,y)=A_\mu (x) \;\theta(n-|x|)\;\; G(x,y)$$
where $G(x,y)\equiv \int d^4 \;p {e^{ip(x-y)}\over(p^4+1)}$
 is the euclidian $4$-dimensional Green function of $\de^4+1$.
It easy to check that the kernels $K_n$ all  satisfy condition (\ref{D_1}),
 so that   $A_n(\de^4+1)^{-1}$
and $A(\de^4+1)^{-1}$ are compact. It means that
$A$ is $\dirac_0^4$-compact and therefore  we can apply the
previous  theorem to $\dirac = \dirac_0 +A$,  obtaining  the
result:
$$\si (\dirac)\equiv \si _{ess} (\dirac)=(-\infi ,+ \infi ).$$
We conclude that if $\dirac$ has eigenvalues (with normalisable
wave functions)  they are embedded into continuous spectrum
and they are not isolated. {\it Q.E.D.}

\section{ Derivation of  Eq.(\protect\ref{3_5}). }

In the basis (\ref{3_1})  write:
$$\dbar=\left[ \matrix{ d & v_n \cr
                        w_m & X_{mn} \cr}\right],$$
$$ \dinv=\left[ \matrix{ d' & v'_m \cr
                        w'_n & X'_{nm} \cr}\right].$$
Imposing $\dbar \dinv=1$ one has (summed indices are left implicit):
\beq d d'+v w'=1 \label{C_2}\eeq
\beq w d'+X w'=0 \label{C_3}\eeq
\beq d v'+v X'=0 \label{C_4}\eeq
\beq w\otimes v'+X X'=1 \label{C_5}.\eeq
Solving these equations for the primed quantities one derives Eq.(\ref{3_7}).\

\noindent
Also $\dbar \dinv=1$ gives:
\beq d' d+v' w=1 \label{C_2'}\eeq
\beq w' d+X' w=0 \label{C_3'}\eeq
\beq d' v+v' X=0 \label{C_4'}\eeq
\beq w'\otimes v+X' X=1 \label{C_5'}.\eeq
Using Eq.(\ref{C_5'}) and then Eq.(\ref{C_3}) one finds:
\beq X'^{-1}-X^{-1}
=-w'\otimes (v X^{-1}) =d'(X^{-1}w)\otimes (vX^{-1}) \label{C_6}.\eeq
Note also the relations
\beq w'=-d' X^{-1}w \label{C_7},\eeq
\beq v'=-d'vX^{-1}\label{C_8} ,\eeq
following respectively from Eq.(\ref{C_3}) and from Eq.(\ref{C_4'}).
Using Eq.(\ref{C_6}-\ref{C_8}) one obtains:
 \bea
 \langle x| \dinv |y \rangle
&=&\, \langle x|a,0 \rangle d' \langle i,0|y \rangle
    + \!\sum_{m\ne 0}\langle x|a,m\rangle w'_m \langle i,0|y \rangle\non\\
&+&\sum_{n\ne 0}\langle x|a,0 \rangle
v'_n \langle i,n|y \rangle
      +\!\sum_{m,n\ne 0}\langle x|a,m\rangle
X'_{mn} \langle i,n|y \rangle  \non \\
&=&\, \langle x|a,0 \rangle d' \langle i,0|y \rangle
    - d'\!\sum_{m\ne 0}\langle x|a,m\rangle (X^{-1}w)_m \langle i,0|y \rangle
     \non\\
&-&d'\sum_{n\ne 0}\langle x|a,0 \rangle
(vX^{-1})_n \langle i,n|y \rangle
      +\!\sum_{m,n\ne 0}\langle x|a,m\rangle
({X'}_{mn}^{-1}-X_{mn}^{-1}) \langle i,n|y \rangle \non  \\
&+&\!\sum_{m,n\ne 0}\langle x|a,m\rangle
X_{mn}^{-1} \langle i,n|y \rangle \} \non \\
&=&d'(\langle x |a,0 \rangle -
\langle x |X^{-1}\bar{C}|a,0 \rangle)
(\langle i,0| y \rangle -
\langle i,0|\bar{B}X^{-1}| y \rangle)
+\langle x|X^{-1}|y \rangle \non \\
&&
\label{C_8bis}\eea
The inverse of a matrix $M$ is $M^{-1}=\det {M} ^{-1}
 \hbox {\rm Cof }
M^{t}$, so that
\beq\det {\dbar}\, d'=\det {X} \label{C_9}.\eeq
 Substituting (\ref{C_9})
 in (\ref{C_8bis}), one derives the expression Eq.(\ref{3_8}) for
$$I(x,y)=\det \dbar  \langle x| \dinv |y \rangle. $$

\section { Perturbative Approach to the Spectrum of $D\dbar$.}

We consider the hermitean 'Hamiltonian' $H=-D\dbar$
 acting in the space
of functions normalizable in a box of size $L\gg \rho, R$
(with periodic boundary conditions). We wish to study the
spectrum of this operator with the ordinary  perturbation theory
of discrete spectrum (in the limit of $L\rightarrow \infty$ the
spectrum of $H$ is continuous, see Appendix C).

For unperturbed hamiltonian we take $H_0=-\de^2$, with eigenfunctions
$$\langle x|k\rangle
={1\over 4L^2} u(k)\, e^{ikx} \quad k={\pi{\bf n} \over L}$$
associated with  eigenvalues $E^{(0)}_k=k^2$. With this choice, the
perturbation
is
$$V=A\bar{\de}+\de \bar{A} -A^2$$
where $A=-ig \si_{\mu} A_{\mu}^{(valley)}$. We take the valley background
in the 'clever' gauge presented in Section 2, in which $A$ is everywhere
regular and behaves at infinity as
\beq A(x)\sim {R\over x^2}\label{E_1}\eeq
(we neglect all spin and isospin factors that are not relevant
 to our approximate analysis).

Our aim is to show that the spectrum of $H$, for $kR\ll 1$ is very close
 to that of $H_0$,
 i.e. that perturbative corrections are small compared to $E^{(0)}_k$
and perturbation theory is well posed.

Consider a generic unperturbed level $E^{(0)}_k$, which is degenerate
(apart the case $k=0$), thus first correction $E^{(1)}_k$ is one of eigenvalues
of the matrix
\bea
V_{kk'}&=&\langle k|V|k'\rangle \\
&\simeq & i\langle k|q_{\mu}\bar{A}_{\mu}+2k_{\mu}A_{\mu}|k'\rangle,
\eea
where we have neglected the second order term $A^2$ and we have set
$q=k-k'$.

 For our purposes it is sufficient to consider
$$V_{kk'}=V(q)\simeq {\max(k,q) \over 16L^4} I(q)$$
where we defined $I(q)\equiv\int^{L} d^4\!  x e^{iqx} A(x)$
(from here on we consider only moduli of vectors).
Note that for degenerate levels $k^2=k^{'2}$ it follows that $0<|q|<2|k|$.

For values of $q$ such that $qL\sim 1$, the
main contribution in the integration comes
from the
region at infinity, where  we can use Eq.(\ref{E_1}) for $A$, arriving at the
estimate
\beq {I(q)\over I(0)}=
{4\over (q L)^2}(1-J_0(qL))\sim {4\over (\pi m)^2}
;\,\,\, q={\pi m\over l}\label{E_2}\eeq
(we used here a spherical box).

The Riemann-Lebesgue theorem guarantees also  that, for a fixed $L$,
$$\lim_{q\rightarrow \infty}\int^{L} d^4\!  x e^{iqx} A(x)=0,$$
which implies  that
\beq\lim_{q\rightarrow \infty}{I(q)\over I(0)}=0\label{E_3}.\eeq
Equations (\ref{E_2}) and (\ref{E_3}) lead  to the relation:
\beq|{V(q)\over V(0)}|\leq {\max (k,q)\over k} ;\quad
 V(0)\equiv({\pi\over 4L})^2 kR
 \label{E_4},\eeq
which we assume to be valid for all $q$.

The degeneracy of each level $k={\pi{\bf n} \over L}$ is of order $M\sim n^3$
($n=|{\bf n}|$),
thus, using Eq.(\ref{E_4}), we can roughly bound the eigenvalues of
perturbation
matrix $V_{kk'}$ by:
\beq|E^{(1)}_{kj}|\leq M^2 |V(0)|\sim {n^7 R \over L^3},\quad j=1,\cdots M
\label{E_5}.\eeq

For the perturbation theory to be good, the first order correction
must be small compared both to the zeroth order value and to the gap
with the nearest nondegenerate energy level:
$$  |E^{(1)}_{k,j}|\ll |E^{(0)}_{k}|,|E^{(0)}_{k+{\pi \over L}}-E^{(0)}_{k}|.$$
 Using the bound Eq.(\ref{E_5}) as an extimate for $|E^{(1)}_{kj}|$, we  see
that
 these  conditions are indeed satisfied for $n^6\ll {L\over R}$,
i.e. substantially for $kR\ll 1$, as we claimed.

Also, from Eq.(\ref{E_5}) it follows that $|V_{k'k}|\ll |k^2-k^{'2}|$ for all
 $k'$,
$$\Delta \psi_{k,j}=\sum_{k'^2\neq k^2}{V_{k'k}\over k^2-k^{'2}}
\psi_{k,j}^{(0)}$$
($\psi_{k,j}^{(0)}$ are eigenfunctions of matrix $V_{k'k}$ and
of course differ
from unperturbed eigenfunctions, due to degeneracy),
substantiating our claim about smallness of corrections to the wave function.

\section { Conventions and Formulae }
In  this Appendix are collected explicit expressions of many quantities
frequently used in the text.

We use the definition of covariant derivative:
$$D^{c}_{\mu }\equiv (\de-igA^{c})_\mu ; \,\,\,
 c=valley,i,a$$
where external fields are given by:
\bea
 A_{\mu}^{(valley)}& =&  A_{\mu}^{(a)} +  A_{\mu}^{(i)} +   A_{\mu}^{(int.
)}, \non\\
A_{\mu}^{(a)}   & =& -{i\over g}(\sigma_{\mu} {\bar \sigma_{\nu}}
-\delta_{\mu \nu} ) {(x-x_a)_{\nu} \over (x-x_a)^2 +\rho^2 },  \non\\
A_{\mu}^{(i)}    &=& -{i\over g}(\sigma_{\mu} {\bar \sigma_{\nu}}
-\delta_{\mu \nu} )
{(x-x_i)_{\nu} \rho^2 \over (x-x_i)^2 ((x-x_i)^2 +\rho^2) },\non\\
A_{\mu}^{(int.)}    &=& -{i\over g}(\sigma_{\mu} {\bar \sigma_{\nu}}
-\delta_{\mu \nu} )
 [ {(x-x_i +y)_{\nu} \over
(x-x_i + y )^2  } -  {(x-x_i)_{\nu} \over
(x-x_i)^2  }],\label{B_1}
\eea
and
$$ y = -R/(z-1);\,\,\,\,
 z = (R^2 + 2 \rho^2 + \sqrt{R^4 + 4\rho^2 R^2}) /2\rho^2;\,\,\,\,
R^{\mu} = (x_i - x_a)^{\mu}.$$
Note that, for ${ R\over \rho}\gg 1$ $|y|\sim {\rho^2\over R}$.

In the text we use the definitions:
$$ D\equiv \si_\mu D^{valley}_\mu ;\,\,\,\dbar \equiv
\sb_\mu D^{valley}_\mu $$
\beq  D^{c}\equiv \si_\mu D^{c}_\mu ; \,\,\,\bar{D^{c}} \equiv
\sb_\mu D^{c}_\mu ;\,\,\, c=i,a\label{B_2}\eeq
where we adopted the convention
$$\si_\mu =(i,\vec{\si}) ;\,\,\,\, \sb_\mu =(-i,\vec{\si}).$$
In spite of the  indistinguished use of the matrices $\si_\mu , \sb_\mu  $,
both for the colour (as in (\ref{B_1})) and spin (as in (\ref{B_2})),
 the reader will easily distinguish them from the context.

The fields $C_\mu,B_\mu$ are defined by
$$D^{valley}_\mu= D^{a}_\mu +C_\mu =D^{i}_\mu +B_\mu$$
so that
$$C_\mu =-ig(A^{(i)}+A^{(int)})_\mu ;\,\,\,B _\mu
=-ig(A^{(a)}+A^{(int)})_\mu.$$

Expressions of zero modes of operators $\dabar$ and $D^{(i)}$ are:
$$
{\dbar}_{j}^{(a)\dot{\alpha}\alpha k}\eta^{(a)}_{0\,\alpha k}(x) =0
;\,\,\,\, \eta^{(a)}_{0\,\alpha k}(x)= -{\rho \over \pi }
{f^{(a)}(x)}^{3/2} \epsilon_{\alpha k},
$$
$${D^{(i)j}}_{\alpha \dot{\alpha} k}
{{\bar \zeta}_{0}}^{(i)\dot{\alpha} k}(x)=0;\,\,\,\,
{{\bar \zeta}_{0}}^{(i)\dot{\alpha} k}(x)={\rho \over \pi }
{f^{(i)}(x)}^{3/2}
{(\si_\mu (x-x_i )_\mu )^{kl} \over \sqrt{(x-x_i )^{2}}}
\epsilon^{l\dot{\alpha}},$$
where we defined
 $$f^{(c)}(x)={1\over (x-x_c )^{2}+\rho^2}$$
 for $ c=i,a$.

Asymptotic behaviour of various quantities is:
$$
\et0(x)\sim {\rho \over (x-x_a)^3} ;\quad \zet0(x) \sim
{\rho \over (x-x_i)^3} $$
$$
C_\mu(x)\sim {\rho \over (x-x_i)^3}+{\rho^2\over R (x-x_i)^2}
$$
$$
B_\mu(x)\sim{1\over x-x_a}+{\rho^2\over R (x-x_i)^2}.$$

However
$$B_\mu-U_a\de_\mu U_a^{\dagger}\sim {\rho^2\over (x-x_a)^3} +
 {\rho^2\over R (x-x_i)^2}$$
with $   U_a = {{\bar \sigma}_{\mu} (x-x_a)_{\mu}
\over \sqrt{(x-x_a)^2}} $
which shows that  the dominant term of $B_\mu$ at infinity has a pure gauge
form.

{\bf Figure Captions}

\begin{description}

\item
{Fig. 1.}  Each flavour $j=3,4,\cdots N_F$ contributes a factor $\rho^2/R^3$.

\item
{Fig. 2.} The propagators for $j=3,4,\cdots N_F$, cut at the intermediate
state.

\item
{Fig. 3.} $C(x_4)$ versus $x_4/\rho$ for ${R\over \rho} =10$ (outmost curve),
${R\over \rho} =5$, ${R\over \rho} =2$ (middle), ${R\over \rho} =1$ and
${R\over \rho} =0.5$ (innermost curve).

\item
{Fig. 4.} C(0) as a function of ${R\over \rho}$.

\item
{Fig. 5.} Valley action ${g^2\over 16 \pi^2} S^{valley}$
as a function of ${R\over \rho}$ (taken from \cite{15}).

\end{description}

\end{document}